\def\a{\alpha}\def\b^2{\lambda}\def\d{\delta}\def\e{\epsilon}
\def\h{\theta}
\def\l{\lambda}\def\m{\mu}\def\n{\nu}\def
\p{\pi}\def\r{\rho}
\def\y{\eta}

\def\O{\Omega}

\def\id{\equiv}\def\mo{{-1}}

\def\({\left(}\def\){\right)}\def\[{\left[}\def\]{\right]}

\def\mn{{\mu\nu}}

\def\pb{Poisson brackets }
\def\sch{Schwarzschild }\def\coo{coordinates }

\def\wrt{with respect to }

\documentclass[showpacs,preprintnumbers,amsmath,amssymb]{revtex4}
\usepackage{graphicx}% Include figure files
\usepackage{float}
\usepackage{dcolumn}% Align table columns on decimal point
\usepackage{bm}% bold math
\begin{document}
\title{Bumblebee gravity and particle motion in Snyder
noncommutative spacetime structures}

% Force line breaks with \\
\author{Sohan Kumar Jha}
\affiliation{Chandernagore College, Chandernagore, Hooghly, West
Bengal, India}
\author{Himangshu Barman, Anisur Rahaman}
 \email{anisur.rahman@saha.ac.in;
 manisurn@gmail.com (Corresponding Author)}
\affiliation{Hooghly Mohsin College, Chinsurah, Hooghly - 712101,
West Bengal, India}

\date{\today}% It is always \today, today,

\begin{abstract}
\begin{center}
Abstract
\end{center}
A metric with a Lorentz violating parameter is associated with the
bumblebee gravity model. We study the motion of a particle in this
bumblebee background where the dynamical variables satisfy
non-canonical Snyder algebra along with some critical survey on
the classical observations in the bumblebee background to see how
these would likely differ from Schwarzschild background. It has
been found that the perihelion shift acquires a generalized
expression with two independent parameters. One of these two is
connected with the Lorentz violating factor and the other is
involved in the Snyder algebraic formulation. We also observe that
the time period of revolution, in general, acquires a Lorentz
violating factor in the bumblebee background, however, for the
circular orbit, it remains unchanged even in the presence of the
Lorentz violating factor in the bumblebee background. The
parameters used here can be constrained with the same type of
conjecture used earlier.
\end{abstract}
%\pacs{11.15Tk, 04.60Kz}
 \maketitle
\section{Introduction}

The standard model of particle physics and the general theory of
relativity are two successful theories that describe the physical
phenomena of nature in two different energy scales and unification
of these two require quantization of the gravity in an essential
manner. Although there are several quite successful and promising
approaches towards quantization of gravity such as string theory
and loop quantum gravity, but these approaches still have some
open issues. Therefore, there lies a deep and increasing interest
in some phenomenological approaches which incorporate some
specific features of the fundamental theories, but that indeed
allow to study the problems typical of some classical frameworks.
One of the open questions which appear due to the development of
various approaches to quantum gravity are the Lorentz symmetry
violation, which might appear in the vicinity of the Planck scale.
However, it is possible that some signals of quantum gravity can
emerge at sufficiently low energy scales. In recent times,
violation of Lorentz symmetry is being taken up as a fascinating
as well as a persuasive concept since string theory \cite{KOST0,
KOSTSAM1, KOSTSAM2, KOSTPOT}, and loop quantum gravity theory
\cite{GAMB, ELLIS} does approve this concept in a captivating
manner. The Lorentz symmetry violation connected to the
electromagnetic background in the interacting field theoretical
model has been extensively studied in the articles \cite{KOSTEM0,
KOSTEM1,KOSTEM2,KOSTEM3, KOSTEM4,KOSTEM5,KOSTEM6, BAKEM,NOVAEM,
RJEM, ADAMEM, FRMEM, SCHEM}. The same effect in connection with
gravity has been studied in the articles
\cite{KOST1,KOST2,KOST3,KOST4, KOST5, KOSTBAIL1, KOSTBAIL2,
KOSTSANTO1, KOSTSANTO2, KOSTCAS}.

Another issue that might appear near the Plank scale is the effect
of the use of non-commutative geometry. So the use of
non-commutative spacetime is also another way of taking the
quantum gravity effect into account. Although this noncommutative
formulation remained ignored for quite a long time, it acquired
huge attention when the noncommutative perspective of spacetime
exhibit its remarkable connection with string theory. Different
physical systems have been studied with this noncommutative
framework in the articles \cite{BI, BIH, BIE, MIG, MIGQFT,
MIGGRAV, MIGPI, SG1, SG2, SG3, HFELD, NOZMDR, BRM, PPMDR, ARMDR,
RABIN}. Gradually modified uncertainty relation has also been
included in this pursuit \cite{DAMITQM, KEMPF}. Although the
validity of noncommutative geometry is presumably limited to
Planck-scale physics like the Lorentz symmetry violation, in the
article \cite{MIGGRAV} it has been shown its effect can be
investigated to macroscopic systems, where the classical limit
holds. In the articles \cite{MIGGRAV, BI} the authors conjectured
that the predicted observation of a microscopic model in a scale
well below the Planck scale would be reasonable to include in
dimensional grounds and some correction is made for a particle in
a Snyder planetary orbits in a relativistic setting.

In the article \cite{CASANA} the authors studied the Lorentz
symmetry violation for the planetary system and in the article
\cite{BI}, the effect of non-commutativity is tested for the same
system. The Lorentz violating (LV) effect is associated with the
interesting model which incorporates spontaneous breaking of
Lorentz symmetry is the so-called bumblebee gravity. On the other
hand, an elegant way to get quantum gravity effect is the
inclusion of noncommutative spacetime structure that admits to a
fundamental length which is the outcome of the string theory and
loop quantum gravity. In general, a noncommutative setting breaks
the Poincare symmetry. However, the pioneering formulation
spacetime non-commutativity due to Snyder \cite{SNY} does admit a
fundamental length scale protecting Lorentz invariance. In fact,
the formulation remains invariant under the celebrated Lorentz
group. This formication has some advantages in our study because
of the underlying symmetry in its formulation.

Now it would be of interest if we get both the effect supposed to
appear at the same energy scale (Planck-scale) for a macroscopic
system from a unified formulation. This article is an endeavor to
formulate it and study both the for a macroscopic system. Since
both the effects may arise in the vicinity of Planck-scale, it
would be reasonable to consider both the effect in a unified
formulation. For this purpose, we need a microscopic system which
is tannable to study both the effects. From the previous works
\cite{CASANA}, it is known that planetary motion can be studied
fruitfully with the LV effect and in \cite{MIGGRAV} the effect due
to the use of Snyder like non-commutative spacetime is studied for
the same system. Therefore, if we are now intended to get the
effect due to the noncommutative spacetime along with the effect
due to the Lorentz symmetry violation offered by the bumblebee
gravity, the noncommutative formulation of spacetime due to Snyder
\cite{SNY} would be operative effectively since it is expected
that the symmetry involved in it would not disturb the effect due
to the Lorentz symmetry violation due to the bumblebee background.
On the other hand, an arbitrary non-commutative setting having no
Lorentz symmetry may have a drastic influence on the LV effect due
to bumblebee. Therefore, this generalization would be tenable for
this system and would certainly be instructive. This
generalization may be thought of as an extension of the work of
Casana {\it etal}. \cite{CASANA} in the noncommutative framework.
As suggested and followed in the articles \cite{MIGGRAV} and
\cite{CASANA} both the correction need to consider in the
dimensional ground since it is a macroscopic system which is in
the energy scale well below the Planck scale.

The article is organized as follows. In Sec. II, we give a brief
introduction to bumblebee gravity. In Sec. III, we have computed
the expression of redshift and surface gravity corresponding to
the bumblebee gravity and found that redshift remains unaltered
but surface gravity does not remain the same. Sec IV. contains a
computation of the time period of a particle revolving in a
circular orbit in the bumblebee background. In Sec. V, we give a
brief introduction of the Snyder algebra which works as an input
to study the motion of a particle in the bumblebee background in
the Snyder framework. In Sec. VI study the motion of a particle in
the bumblebee background where the canonical variable satisfies
Snyder algebra, and the final Sec. VII contains a brief summary
and discussion.
%%%%%%%%%%%%%%%%%%%%%%%%%%%%%%%%%%%%%%%%%%%%%%%%%%%
\section{Bumblebee Model}
It is model where Lorentz symmetry breaking is manifested through
a bumblebee field vector $B_\mu$. The under lying idea is simple
however it shows interesting features like CPT violation along
with the spontaneous breaking of Lorentz symmetry. This
interesting model is  described by the action
\begin{equation}
S_{BB}=\int d^4x[\frac{1}{16\pi G_N}({\cal R}+\eta
B^{\mu}B^{\nu}{\cal R}_{\mu\nu})]-\frac{1}{4}{\cal
B}_{\mu\nu}B^{\mu\nu}-V(B_\mu B^\mu \pm b_2)]. \label{BA}
\end{equation}
Here $\eta$ has the dimension of $M^{-2}$. The Ricci tensor is
represented by $R_{\mu\nu}$ and the bumblebee field strength is
defined by
\begin{equation}
B_{\mu\nu} = \nabla_\mu B_\nu - \nabla_\nu B_\mu
\end{equation}
$V$ represents  the potential  defined through the bumblebee field
that navigate the breaking  of Lorentz symmetry for the system
which collapses onto a nonzero minima with the condition $B_\mu
B^\mu = \pm b_2$. The Lagrangian density of the bumblebee gravity
model \cite{BB0, BB1, BB2} results the following extended vacuum
Einstein equations
\begin{eqnarray}
G_{\mu\nu}=R_{\mu\nu}-\frac{1}{2}Rg_{\mu\nu}=\kappa T_{\mu\nu}^{B}
\end{eqnarray}
Here  $G_{\mu\nu}$ and $T_{\mu\nu}^{B}$  are respectively
representing Einstein and bumblebee energy-momentum tensors. The
gravitational coupling  ${\kappa=8\pi G}_{{N}}$ and  the bumblebee
energy momentum tensor $T_{\mu\nu}^{B}$ has the following
expression
\begin{eqnarray}
T_{\mu\nu}^{B} & =
&-B_{\mu\alpha}B_{~\nu}^{\alpha}-\frac{1}{4}B_{\alpha\beta
}B^{\alpha\beta}g_{\mu\nu}-Vg_{\mu\nu}+2V^{\prime}B_{\mu}B_{\nu}+\frac{\xi
}{\kappa}\left[
\frac{1}{2}B^{\alpha}B^{\beta}R_{\alpha\beta}g_{\mu\nu
}-B_{\mu}B^{\alpha}R_{\alpha\nu}\right.\nonumber  \\
& - & B_{\nu}B^{\alpha}R_{\alpha\mu}+\frac{1}{2}\nabla_{\alpha}
\nabla_{\mu}(  B^{\alpha}B_{\nu})  +\frac{1}{2}\nabla_{\alpha
}\nabla_{\nu}(  B^{\alpha}B_{\mu})  -\frac{1}{2}\nabla^{2}(
B_{\mu}B_{\nu}) -\frac{1}{2}g_{\mu\nu}\nabla_{\alpha}\nabla_{\beta
}( B^{\alpha}B^{\beta})].
\end{eqnarray}
The dimensional coupling constant $\xi$ having dimension $M^{-1}$
is representing the real coupling constant in this situation.
Using the contracted Bianchi identities after taking the covariant
divergence of the Einstein-bumblebee equation one obtains
\begin{equation}
\nabla^{\mu}T_{\mu\nu}^{B}=0.
\end{equation}
It is the  covariant conservation law for the energy-momentum
tensor. This helps to land on to
\begin{equation}
R_{\mu\nu}=\kappa
T_{\mu\nu}^{B}+\frac{\xi}{4}g_{\mu\nu}\nabla^{2}\left(
B_{\alpha}B^{\alpha}\right)  +\frac{\xi}{2}g_{\mu\nu}\nabla_{\alpha}%
\nabla_{\beta}(B^{\alpha}B^{\beta}).
\end{equation}
One can immediately see that when the bumblebee field $B_{\mu}$
vanishes, we recover the ordinary Einstein equations. When the
bumblebee field remains frozen in its vacuum expectation value
(VEV) we are allowed to write
\begin{equation}
B_{\mu}=b_{\mu}\qquad
\Rightarrow{b}_{\mu\nu}\equiv\partial_{\mu}b_{\nu}-\partial_{\nu}
b_{\mu}.
\end{equation}
In this situation,  Einstein equations acquires a generalized form
\begin{eqnarray}
& & R_{\mu\nu}+\kappa
b_{\mu\alpha}b_{~\,\nu}^{\alpha}+\frac{\kappa}
{4}b_{\alpha\beta}b^{\alpha\beta}g_{\mu\nu}+\xi
b_{\mu}b^{\alpha}R_{\alpha\nu }+\xi
b_{\nu}b^{\alpha}R_{\alpha\mu}-\frac{\xi}{2}b^{\alpha}b^{\beta}
R_{\alpha\beta}g_{\mu\nu}\nonumber\\
&-&\frac{\xi}{2}\nabla_{\alpha}\nabla_{\mu}(  b^{\alpha}b_{\nu})
-\frac{\xi}{2}\nabla_{\alpha}\nabla_{\nu}(  b^{\alpha}b_{\mu})
+\frac{\xi}{2}\nabla^{2}\left(  b_{\mu}b_{\nu}\right)  = 0.
\label{GEE}
\end{eqnarray}
If $b_{\mu}=[0,b_{r}(r),0,0]$ is set as space like background for
the bumblebee field $b_{\mu}$ which satisfies the condition
$b^{\mu}b_{\mu}=b^{2}=$\textit{constant} a  spherically symmetric
static vacuum solution to equation (\ref{GEE}) results in:
\begin{equation}
ds^{2}=-\left(  1-\frac{2M}{r}\right)  dt^{2}+\left(  1+l\right)
\left( 1-\frac{2M}{r}\right)  ^{-1}dr^{2}+r^{2}\left(
d\theta^{2}+\sin^{2}\theta d\varphi^{2}\right),
\end{equation}
where the parameter $l$ is defined by $l=\xi b^{2}\geq 0$ is
termed as  Lorentz symmetry breaking (LSB) parameter. Although the
metric has a structural similarity with the Schwarzschild metric
it is known to be very different from that since  this metric
renders the following  Kretschmann scalar
\begin{equation} \mathcal{K}=\frac{4\left(
12M^{2}+4LMr+l^{2}r^{2}\right) }{r^{6}\left( 1+l\right)  ^{2}}.
\end{equation}
As a consequence  the Hawking temperature corresponding to the
metric  takes a $l$ dependent generalized form:
\begin{equation}
T_{H}=\frac{1}{4\pi\sqrt{-g_{tt}g_{rr}}}\left.
\frac{dg_{tt}}{dr}\right\vert _{r=r_{h}}=\left.
\frac{1}{2\pi\sqrt{1+l}}\frac{M}{r^{2}}\right\vert
_{r=r_{h}}=\frac{1}{8\pi M\sqrt{1+l}}
\end{equation}.
These are all known about the bumblebee gravity. To what extent
the classical dynamics in bumblebee gravity differs from the
Schwarzschild would be instructive which we would like to
investigate to start with. We should admit that some consequences
of bumblebee gravity background is known still some consequences
of bumblebee background are still there which should  be brought
to light to which we now turn.

%%%%%%%%%%%%%%%%%%%%%%%%%%%%%%%%%%%%%%%%%%%%%%%%%%%%%
\section{Redshift and Surface Gravity for bumblebee gravity}
Before delving into our main problem we study some classical tests
related to bumblebee gravity which are well known for
Schwarzschild metric.
\subsection{Redshift}
Let us first consider the redshift scenario in the bumblebee
gravity. For bumblebee background we have the same expression of
proper time as we find in Schwarzschild background.
\begin{equation}
d{\tau}=(1-\frac{2M}{r})^{\frac{1}{2}}dt.
\end{equation}
It is known that  the difference in the measurement of time
locally and at infinity is the redshift, since the wavelength of
radiation is proportional to the period of vibration, equation
$(12)$ gives us
\begin{equation}
\tilde{\lambda}=(1-\frac{2M}{r})^{\frac{1}{2}}{\tilde{\lambda}}_{\infty}
\end{equation}
for the relation between the wavelength $\lambda$ of radiation
emitted at r and the wavelength $\tilde{\lambda}_{\infty}$
received at infinity. For the redshift $z$, defined by
\begin{equation}
z=\frac{\tilde{\lambda}_{\infty}-\tilde{\lambda}}{\tilde{\lambda}}
\end{equation}
we get
\begin{equation}
1+z=(1-\frac{2M}{r})^{-\frac{1}{2}}
\end{equation}
which is the same as that for \sch background. Thus amount of
redshift does not depend on LV factor $l$. However the other
features does not remain the same which will be revealed in the
following sections.

\subsection{Surface Gravity}
We now proceed to calculate the surface gravity in this new
background. It is known that surface gravity is the local proper
acceleration multiplied by the gravitational time dilation factor.
From the standard definition the proper acceleration is given by
\begin{equation}
a^{\mu}a_{\mu}=-a^2
\end{equation}
where 4-acceleration $a^{\mu}$ is given by
\begin{equation}
a^{\mu}=\frac{du^{\mu}}{d{\tau}}+{\Gamma}_{\rho\sigma}^{\mu}u^{\rho}u^{\sigma}.
\end{equation}
Let us now write down  the component of the 4-velocity of a
hovering observer:
\begin{equation}
u^{\mu}=((1-\frac{2M}{r})^{-\frac{1}{2}},0,0,0).
\end{equation}
The only non-zero component that the 4-acceleration contains is
\begin{equation}
a^1=\frac{du^1}{d{\tau}}+{\Gamma}_{00}^1(u^0)^2,
\end{equation}
where ${\Gamma}_{00}^1=\frac{M}{r^2(1+l)}(1-\frac{2M}{r})$.  Since
$u^1=0$, we obtain
\begin{equation}
a^1=\frac{M}{r^2(1+l)},\qquad a_1=g_{11}a^1.
\end{equation}
We therefore have
\begin{equation}
a^{\mu}a_{\mu}=-\frac{1}{1+l}(\frac{M}{r^2})^2(1-\frac{2M}{r})^{-1},
\end{equation}
which ultimately gives
$a=\frac{1}{\sqrt{1+l}}\frac{M}{r^2}(1-\frac{2M}{r})^{-\frac{1}{2}}$.
The time dilation factor in this situation is
$(1-\frac{2M}{r})^{\frac{1}{2}}$. Thus the  expression of surface
gravity  turns into
\begin{eqnarray}
\chi_{BB}= \frac{M}{r^2\sqrt{1+l}}.
 \end{eqnarray}
For Schwarzschild radius $r=2GM$ that becomes
$\frac{1}{4M\sqrt{1+l}}$ . It shows that the surface gravity for
the bumblebee background depends on the LV factor $l$ in an
explicit manner. While the redshift scenario does remain
unaffected by the LV factor the surface gravity does not remain
so. One more surprise is laid in the study of particle motion
under this background to which we now turn.
\section{Circular orbit in bumblebee background}
It is known that the perihelion shift acquires an LV factor for
general orbit but does it hold for the motion of circular orbit?
To see this let us study the motion of a particle in a circular
orbit in the bumblebee background as an independent problem. For
bumblebee background we have
\begin{equation}
1=(1-\frac{2M}{r})(\frac{dt}{d\tau})^2-(1+l)(1-\frac{2M}{r})^{-1}(\frac{dr}{d\tau})^2
-r^2(\frac{d\phi}{d\tau})^2.
\end{equation}
Along any geodesic in the presence of bumblebee background we can,
therefore, obtain
\begin{equation}
(1-\frac{2M}{r})\frac{dt}{d\tau}=E=constant.
\end{equation}
Here $E$ is the relativistic energy per unit mass of the particle
relative to a stationary observer at infinity. We have one more
information regarding this motion
\begin{equation}
r^2\frac{d\phi}{d\tau}=L=constant,
\end{equation}
where $L$ is the angular momentum per unit mass relative to an
observer at infinity. Using the above two equations we obtain
\begin{equation}
(\frac{dr}{d\tau})^2=\frac{E^2}{1+l}-\frac{1}{1+l}(1+\frac{L^2}{r^2})(1-\frac{2M}{r}).
\end{equation}
For a circular orbit the following two conditions
\begin{equation}
\frac{dr}{d\tau}=0\qquad\qquad\qquad\qquad \frac{d^2r}{d\tau^2}=0,
\end{equation}
must be satisfied. The first condition gives gives
\begin{equation}
E=\sqrt{(1+\frac{L^2}{r^2})(1-\frac{2M}{r})}, \label{EE}
\end{equation}
and the second condition after some simplification turns into
\begin{equation}
Mr^2-L^2r+3ML^2=0. \label{RR}
\end{equation}
The quadratic equation (\ref{RR}) yields the following solution
for $r$
\begin{equation}
r=\frac{L^2}{2M}\pm \frac{1}{2}\sqrt{\frac{L^4}{M^2}-12L^2}.
\end{equation}
With this $r$ the expression of  $L$ comes out to be
\begin{equation}
L=\sqrt{\frac{Mr}{1-3Mr}}. \label{EL}
\end{equation}
We get the following expression of $E$ using  equations (\ref{EE})
and (\ref{EL}).
\begin{equation}
E=(1-\frac{2M}{r})(1-\frac{3M}{r})^{-\frac{1}{2}}. \label{ELVI}
\end{equation}
Note that for circular orbit the equation (\ref{ELVI}) has no LV
factor but the energy of a particle in non-circular orbit does not
remain independent of LV factor as it is known from the article
\cite{CASANA}. The above information leads us to compute the time
period of revolution in a circular path as follows.
\begin{equation}
\frac{d\tau}{d\phi}=\frac{r^2}{L}=\frac{r^{\frac{3}{2}}}{\sqrt
M}\sqrt{1-\frac{3M}{r}}.
\end{equation}
Thus proper period of  the particle in the circular orbit is found
out to be
\begin{equation}
\tau=\frac{r^2}{L}=2\pi\frac{r^{\frac{3}{2}}}{\sqrt
M}\sqrt{1-\frac{3M}{r}}.
\end{equation}
We observe that the proper period of a circular orbit in the
bumblebee background is identical with the time period for \sch
background and it does not depend on LV factor $l$. If we look at
the expression of a general non-circular orbit we find that the
time period in that situation for bumblebee background was
exclusively dependent on the LV factor. So the LV factor has no
influence on the symmetric circular orbit! We would like to
mention that this result may follow from the general discussion of
the article \cite{CASANA}, but it was not clearly pointed out
there.

What follows next is much involved: the study of particle motion
in the bumblebee gravity where the canonical variable satisfies
noncommutative but Lorentz Symmetric Snyder algebra. This problem
in the Schwarzschild background was studied in \cite{MIGGRAV}. The
motivation of generalization of it in bumblebee gravity is,
therefore, automatic after the work of Mignemi {\it et al}.
\cite{MIGGRAV}. Besides both the Lorentz effect and the effect due
to the noncommutative setting of originates are suppose to appear
in the vicinity of the Planck scale so the study of the unified
effect is instructive. Before going to study the motion of a
particle in the bumblebee gravity with Snyder noncommutative
framework it will be beneficial to give a brief discussion of
Snyder formulation:

\section{A brief discussion of Snyder algebra}
For this article, a general discussion of Snyder algebra will also
be useful like the introduction of bumblebee gravity given in Sec.
II to make this article self-contained. So before jumping onto the
computation related to particle orbit in the bumblebee background
framework developed by Snyder we give a brief discussion over
general Snyder algebra. The non-canonical generalization of the
canonical Poisson brackets that follow from \cite{SNY} is
\begin{eqnarray}
\{x_\m,p_\n\}=\y_\mn+\lambda p_\m p_\n + \alpha x_\m x_n +
\alpha\lambda p_\m x_\nu ,\qquad\{x_\m,x_\n\}=\lambda
J_\mn,\qquad\{p_\m,p_\n\}=0, \label{GSA}
\end{eqnarray}
where $J_{\mn}=x_\n p_\n-x_\n p_\m$, $\y_\mn$ is the flat metric
with signature $(-1,1,1,1)$ $\a$ and $\lambda$ and $\alpha$ are
two coupling constants and the square of these two are having the
dimension of inverse Planck length and inverse Planck mass
respectively. For $\alpha \rightarrow 0$ the algebra (\ref{GSA})
reduces to the ordinary Snyder non-relativistic model with which
we are interested in our present article. So precisely the algebra
reads

\begin{eqnarray}
\{x_\m,p_\n\}=\y_\mn+\lambda p_\m p_\n
,\qquad\{x_\m,x_\n\}=\lambda J_\mn,\qquad\{p_\m,p_\n\}=0.
\label{OSA}
\end{eqnarray}
In ordinary units, thiscorresponds to
$\lambda^2\sim\sqrt\hbar/cM_{Pl}\sim10^{-17}$(s/kg)$^{1/2}$. The
\pb (\ref{OSA}) preserve the Lorentz invariance. In order to study
the free particle with a bumblebee gravity background with the
non-canonical framework of Snyder, it would be beneficial to work
with Polar coordinate. So the spatial sections with polar
coordinates are parameterized in terms of cartesian \coo as
follows
\begin{eqnarray}
t=x_0=-x^0,\qquad\r=\sqrt{(x^1)^2+(x^2)^2},\qquad\h=tan^{-1}\frac{x^2}
{x^1}.
\end{eqnarray}
Therefore, the momenta corresponding to the polar coordinates can
be written down as
\begin{eqnarray}
p_t=p_0,\qquad p_\r=\frac{x^1p_1+x^2p_2}{\sqrt{(x^1)^2+(x^2)^2}},
\qquad p_\h\id J_{12}=x_1p_2-x_2p_1,
\end{eqnarray}
in order to satisfy the canonical Poisson bracket. Using the
Snyder algebra (\ref{OSA}) it is straightforward to see that the
phase space in the polar coordinate in the Snyder space lose the
symplectic structure and acquire the following modified form
\begin{eqnarray}
\{t,p_t\}&=&-1+\lambda
p_t^2,\qquad\{\r,p_\r\}=1+\lambda\left(p_\r^2+\frac{p_\h^2}{\r^2}\right),
\qquad\{\h,p_\h\}=1, \nonumber
\end{eqnarray}
\begin{eqnarray}
\{\r,\h\}&=&\lambda\frac{p_\h}{\r},\qquad\{t,\r\}=\lambda(tp_\r-\r
p_t), \qquad\{t,\h\}=\lambda\frac{tp_\h}{\r^2},\nonumber
\end{eqnarray}
\begin{eqnarray}\nonumber
\{p_t,p_\r\}&=&-\lambda\frac{p_tp_\h^2}{\r^3},\qquad\{p_t,p_\h\}=\{p_\r,p_\h\}=\{t,p_\h\}=
\{\r,p_\h\}=0,
\end{eqnarray}
\begin{eqnarray}
\{t,p_\r\}&=&\lambda\(p_tp_\r+\frac{tp_\h^2}{\r^3}\),\qquad\{\r,p_t\}=\lambda
 p_tp_\r,\nonumber
\end{eqnarray}
\begin{eqnarray}
\{\h,p_t\}=\lambda \frac{p_t
p_\h}{\r^2},\qquad\{\h,p_\r\}&=&\lambda\frac{p_\r p_\h}{\r^2}.
\end{eqnarray}
The above information will be useful to study the motion of a
particle in a bumblebee background.

\section{Motion of Particle in bumblebee background in the Snyder orbit}
The motion of Particle in bumblebee background in the Snyder orbit
is an extension of the work done in the article \cite{CASANA} with
the inclusion of noncommutative spacetime through the Snyder
formalism. It can also be thought of as an extension of the work
presented in the article \cite{MIGGRAV} where bumblebee background
replaces the Schwarzschild background. The necessary inputs for
this extension are all given in the previous sections. Let us now
proceed to study the motion of a planet (particle) with the metric
consistent with the bumblebee gravity model. Precisely, the metric
is
\begin{eqnarray}
ds^2=-f(\r)\,dt^2+(1+l)f^\mo(\r)\,d\r^2+\r^2d\O^2,
\end{eqnarray}
where $f(\r)=1-\frac{2M}{\r}$ and $M$ is the mass of the Sun. It
is a system defined in $(3+1)$ dimension, however, the azimuthal
symmetry leads to the conservation angular momentum that offers a
simplification which helps to reduce it into a problem of $(2+1)$
dimension like its special relativistic counterpart.

The dynamics of a massive body with mass $m$ in the presence of
bumblebee gravity background is given by the Hamiltonian
\begin{eqnarray}
H=\frac{k}{2}[-\frac{p_t^2}{
f(\r)}+\frac{f(\r)}{1+l}p_\r^2+\frac{p_\h^2}{\r^2}+m^2]=0.
\label{HAM}
\end{eqnarray}
Now we are in a position to study the perihelion shift of a planet
of mass $m$ under the active influence of the bumblebee gravity
with Snyder like Lorentz invariant noncommutative framework. The
equations of motion of the planet under investigation with Snyder
like noncommutative framework under active of presence bumblebee
background can be calculated using the set of equations
(\ref{HAM}) in a straightforward manner.
\begin{eqnarray}
\dot t=k[p_t (A^\mo- \b^2
m^2-\b^2\frac{M}{\r}(p_\r^2+\frac{p_t^2}{ A^2}))+
\b^2\frac{Mtp_\r}{\r^2}(p_\r^2+\frac{p_t^2}{
A^2}-2\frac{p_\h^2}{\r^2(1+l)})-\b^2\frac{p_\h^2ltp_\r}{ \r^3}],
\end{eqnarray}
\begin{eqnarray}
\dot\r =k[\frac{A}{
1+l}-\b^2m^2-\frac{2\b^2Mp_\h^2}{\r^3(1+l)}-\b^2\frac{p_\h^2l}{
\r^2(1+l)}]p_\r,
\end{eqnarray}
\begin{eqnarray}
\dot\h=k\frac{p_\h}{\r^2}[1-\b^2m^2-\frac{\b^2M}{\r}\,\(\frac{p_\r^2}{
1+l}+\frac{p_t^2}{A^2}\)],
\end{eqnarray}
\begin{eqnarray}
\dot p_t=-k[\frac{\b^2Mp_tp_\r}{\r^2}(\frac{p_\r^2}{
1+l}-\frac{2p_\h^2}{\r^2(1+l)}+\frac{p_t^2}{
A^2}-\frac{p_\h^2l}{\r^2M(1+l)})], \qquad\qquad\dot p_\h=0,
\label{EPT}
\end{eqnarray}
\begin{eqnarray}
\dot
p_\r=k[(1-\b^2m^2)\frac{p_\h^2}{r^3}+\frac{\b^2p_\h^2p_\r^2l}{
\r^3(1+l)} -\frac{M}{\r^2}((\frac{p_\r^2}{1+l}+\frac{p_t^2}{ A^2})
(1+\b^2(p_\r^2+\frac{p_\h^2}{\r^2}))-2\b^2\frac{p_\r^2p_\h^2}{\r^2})].
\label{EPP}
\end{eqnarray}
To find out the perihelion shift with this altered scenario is our
principal objective. To get it, we indeed require the equation of
the orbits. Note that $p_t$ is no longer conserved but $p_\h$ is.
So if we look at the expression
\begin{eqnarray}
E=\frac{p_t}{\sqrt{1+\b^2(-p_t^2+p_\r^2+p_\h^2/\r^2)}},\label{FEE}
\end{eqnarray}
we find that it is a conserved quantity, and it denotes the energy
of the particle revolving under bumblebee gravity. Equation
(\ref{EPT}) with the use of the equations (\ref{FEE}) gives
\begin{eqnarray}
p_t^2=\frac{E^2}{1+\b^2E^2}[1+\b^2(p_\r^2+p_\h^2/\r^2)].
\label{PT2}
\end{eqnarray}
The questions (\ref{EPP}) and (\ref{FEE}) results
\begin{eqnarray}
p_\r^2=(1+l)\frac{E^2(1+\b^2m^2h^2/\r^2)-m^2(1+\b^2E^2)(1+h^2/\r^2)A}{(1+\b^2E^2)A^2
-\b^2E^2(1+l)}, \label{PR2}
\end{eqnarray}
where  $h=p_\h/m$ is set for convenience.

It would be beneficial to get the equation of the orbits in terms
of the variable $u$ which is defined by $u=1/\r$. In terms of $u$
the equation looks
\begin{eqnarray}
\frac{du}{d\h}=-\frac{1}{\r^2}\frac{\dot\r}{\dot\h}
=-\frac{\frac{A}{1+l}-\b^2m^2(1+\frac{2Mh^2u^3}{1+l}
+\frac{h^2u^2l}{1+l})}{1-\b^2m^2-\b^2Mu (\frac{p_\r^2}{1+l}
+\frac{p_t^2}{ A^2})}\frac{p_\r}{mh}. \label{DEU}
\end{eqnarray}
If we use the expression for $p_\r$ and $p_t$ from equations
(\ref{PT2}) and (\ref{PR2}) respectively in equation (\ref{DEU})
it will enable us to get a decoupled equation for the single
dimensionless variable $v$, which is defined as $v=\frac{h^2}{
M}\,u$. It will also be helpful if we define another dimensionless
parameter $\e=\frac{M^2}{h^2}$. The parameter $\e$ is small for
planetary orbits, and it can be taken as an expansion parameter.
We also assume that $\b^2m^2\ll\e$ since the Snyder corrections
are expected to be small \wrt the correction appears from general
relativity. Besides, by the virial theorem, and the definition of
$E$ standing in equation (\ref{FEE}) reveals that ($E^2-m^2\sim
m^2~(\e\,q+\b^2E^2))$, where $q$ is a parameter of order unity.
After a bit of algebra with the first-order expansion of both
$\b^2m^2$ and $\e$ results the following first-order second-degree
equation for $v$
\begin{eqnarray}
v'^2=&&\frac{q}{1+l}+\frac{2v}{1+l} -\frac{v^2}{1+l}+\frac{2\e
v^3}{ 1+l}+\b^2m^2\big[-ql\frac{1+2l}{1+l}+\frac{2v}{1+l}
-lv\frac{2+4l}{1+l}\nonumber \\
&+&lv^2\frac{1+2l}{1+l}+\e(\frac{q^2l}{1+l}+qv\frac{4-2l}{
1+l}-qlv\frac{6+4l}{1+l}+v^2\frac{(4-4l)}{1+l}
-qlv^2\frac{3+2l}{1+l}\nonumber\\
&-&8lv^2\frac{2+l}{1+l}+lv^3\frac{4-4l}{1+l}+2lv^4)\big]
\label{SDV}
\end{eqnarray}
To obtain a solution of the second degree equation it is
convenient to take the derivative of equation (\ref{SDV}) to
convert it into a second order differential equation:
\begin{eqnarray}\nonumber
v''=&&\frac{1-\b^2m^2l-2\b^2m^2l^2+\b^2m^2}{1+l}+\frac{\b^2m^2l+2\b^2m^2l^2-1}{1+l}v\nonumber\\
&+&\e(k_0+k_1v+k_2v^2+k_3v^3)
\end{eqnarray}
where $k_0=\b^2m^2\frac{-4ql+2q-2ql^2}{1+l},$
$k_1=\b^2m^2\frac{-3ql-2ql^2-20l-8l^2+4}{1+l},$ $k_2=
\frac{3+6\b^2m^2l-6\b^2m^2l^2}{ 1+l},$ and $k_3=\b^2m^2l$.
 If an
expansion of $v$ up to first order in $\e$ is made as $v=v_0+\e
v_1+\dots$, where
\begin{eqnarray}\nonumber
v_0=1+\b^2m^2+e\cos \gamma\h,\qquad
e=1+\frac{q}{\e}=1+\frac{h^2(E^2-m^2)}{M^2m^2 },
\end{eqnarray}
then the zeroth order, i.e., $v_0$ will render the result that
corresponds to the Newtonian limit as usual, where $v_1$ will
satisfy the differential equation
\begin{eqnarray}
v_1''+\gamma^2v_1=k_0+k_1v_0+k_2v_0^2+k_3v_0^3.
\end{eqnarray}
where
\begin{eqnarray}
\gamma^2=\frac{1}{1+l}-\b^2m^2l\frac{1+2l}{1+l}.
\end{eqnarray}
 The solution of this second-order differential equation  is
found out to be
\begin{eqnarray}
v_1&=&\frac{1}{
\gamma^2}\big[k_0+k_1(1+\b^2m^2)+k_2(1+2\b^2m^2+\frac{e^2}{
2})+k_3(1+3\b^2m^2+\frac{3e^2}{2})\big]+e[k_1\nonumber \\
&+&{2k_2(1+\b^2m^2)}+k_3(\frac{3e^2}{ 4}+3e)]\frac{\h \sin
\gamma\h}{
2w}-[k_2\nonumber\\
&+&3k_3]\frac{e^2}{ 6\gamma^2}\cos 2\gamma\h-\frac{k_3e^3}{
32\gamma^2}\cos 3\gamma\h \label{SOLN}
\end{eqnarray}
In the solution (\ref{SOLN}), the terms that oscillate around zero
are neglected, and hence with the lowest  order it reads
\begin{eqnarray}\nonumber
v&\sim&(1+\b^2m^2)+\frac{\e\,}{
\gamma^2}[k_0+k_1(1+\b^2m^2)+k_2(1+2\b^2m^2+\frac{e^2}{
2})+k_3(1+3\b^2m^2+\frac{3e^2}{2})]\\\nonumber
&+&e\cos [1-\frac{\e\,}{ \gamma^2}(\frac{k_1}{ 2}+k_2(1+\b^2m^2)
+\frac{k_3}{2}(\frac{3e^2}{ 4}+3e))]\gamma\h \nonumber\\
&=&(1+\b^2m^2)+\frac{\e\,}{
\gamma^2}[k_0+k_1(1+\b^2m^2)+k_2(1+2\b^2m^2+\frac{e^2}{
2})+k_3(1+3\b^2m^2+\frac{3e^2}{ 2})]\\\nonumber &+&e\cos\omega\h
\end{eqnarray}
where $\omega= [1-\frac{\e\,}{
\gamma^2}(\frac{k_1}{2}+k_2(1+\b^2m^2)+\frac{k_3}{2}(\frac{3e^2}{
4}+3e))]\gamma$.
 From
the above expression one obtains the generalized expression for
the perihelion shift:
\begin{eqnarray}
\d\h&=&\p\e\,[6+10\b^2m^2+(-3q+1+\frac{3e^2}{4}+3e)\b^2m^2l]+\p\b^2m^2l+5l\p\e\b^2m^2+l\p\nonumber
\\ &=& l\p + 6\p\e +10\p\e\b^2m^2+\p\e(-3q+1+\frac{3e^2}{4}+3e) +
\p(1+ 5\e)\b^2m^2l.
\end{eqnarray}
Note that the expression of time period agrees with the time
period obtained in the article \cite{CASANA} since if we set
$\lambda =0$, $\omega$ will reduces to $\omega\approx
\frac{1-\epsilon}{\sqrt{1+l}}$. The first term represents the
shift due to the curvature of the spacetime. The term containing
only $\lambda$ appears due to the use of Snyder noncommutative
dynamics and the term containing only $l$ emerges because of the
replacement of the Schwarzschild background by the bumblebee.
These two can be considered as pure corrections due to the use of
noncommutative dynamics and the use of bumblebee gravity
respectively. We are inserting hare a table (TABLE 1) containing
the upper bound of the individual correction paraments of each
effect.
\begin{table}[ht!]
  \begin{center}
    \caption{Estimates of upper bounds for various planets}
    \label{tab:table1}
    \begin{tabular}{p{1.1 cm}|c|c|p{3.2 cm}|c|c}
      \textbf{Planet}&\textbf{GR prediction}&\textbf{Observed}&\textbf{Uncertainty arcseconds per century}
      & \textbf{LV  Parameter} & \textbf{Snyder parameter}\\
      \hline
 Mercury & ${42.9814}$ & $\,{42.9794}\,\pm \,0.0030\,$ &$%
-0.0020\,\pm \,0.0030$ &$1.1\times 10^{-11} $ &  $2.0\times 10^{-89} $ \\
      Venus &$8.6247$ &$\,\,8.6273\,\pm
\,0.0016$&   $0.0026\,\pm \,0.0016$ &$1.5\times 10^{-11} $  &  $2.2\times 10^{-91} $ \\
      Earth&  $3.83877$ & \ $\,3.83896\,\pm
\,0.00019$ & $0.00019\,\pm \,0.00019$ &$2.9\times 10^{-12} $ &  $4.3\times 10^{-92} $ \\
      Mars&${1.350938}$ & \ $\,{1.350918}\,\pm
\,0.000037$ & $-0.000020\,\pm \,0.000037$ &$1.1\times 10^{-12} $ &  $2.0\times 10^{-90} $ \\
 Jupiter& $0.0623$ & $%
\,0.121\,\pm \,0.0283$&$0.0587\,\pm \,0.0283$&$5.2\times 10^{-9}$ & $3.7\times 10^{-93} $ \\
Saturn&$ 0.01370$ & \ $\,0.01338\,\pm \,0.00047$ & $%
-0.00032\,\pm \,0.00047$ & $2.1\times 10^{-10} $ & $3.2\times 10^{-93} $ \\
\hline% <-- added row here
    \end{tabular}
  \end{center}
\end{table}
The upper bound if $\lambda$ is in agreement with the prediction
made in \cite{BI}, where it was predicted that it should be less
than $10^{-18}$ for the Mercury. Of course, few terms are there
which maybe considered as an interfering (mixed) term containing
the product of $\lambda l$. These terms are indeed small compared
to the pure term but for this setting, we can not avoid the
presence of it. It has some interesting consequences too. If we
would like to have the contribution of the order of the first term
as it has been conjectured in \cite{MIGGRAV} and \cite{CASANA} to
constrain $\lambda$ and $l$ it will impose a new constraint on
$\lambda$ which shows that these two corrections have an inverse
effect with each other so far magnitude of the correction is
concerned. The enhancement of one will be demised by the other
since the product is put under constraint. Note that if we would
like to constrain $\lambda$ with the consideration that
$\pi\lambda\m^2\l$ will give the same type of contribution as the
other two pure terms then the the parameter $\lambda$ will come
out approximately to $9.4\times10^{-48}$ for Mercury. From the
interfering terms, a single term $\pi\lambda\m^2\l$ is considered
here for computation to get the approximated vale since the order
of magnitude will not change if all the interfering terms are
considered. It can be constrained for other planets as well in
this manner. It also agrees with the prediction made in the
article \cite{MIGGRAV}.
\section{Summary and Discussion}
In this article, initially, we have considered a few classical
tests related to bumblebee gravity. We have seen that although the
redshift scenario remains unaltered, the surface gravity acquires
an LV correction in this background. When we study the motion of a
particle confined in a circular orbit we notice that it also
remains unaltered whereas the motion in a generalized non-circular
orbit does not remain so. The precise investigation on the
circular orbit extracts out this surprising exposition. We would
like to mention here that this result can be followed from the
article \cite{CASANA}, but without going through a separate
calculation this information may not be uncovered. Why it is so
can not be explained at this stage. It may need further
investigation. It may be the case that the increased symmetry in
the circular orbit may protect itself to suffer from the Lorentz
symmetry violation!

Our main endeavor in this article is to study the motion of a
particle in the bumblebee background that contains an LV parameter
with a noncommutative but Lorentz invariant setting formulated by
Snyder. The way we have formulated our investigation is capable of
providing corrections on the perihelion shift of a particle likely
to appear in the vicinity of Planck-scale for two different means
in a unified manner. The replacement of Schwarzschild background
by the bumblebee renders a correction through the LV term $l$
whereas the noncommutative setting of Snyder renders a correction
through the parameter $\lambda$ in the same energy scale. Our
result shows that the correction in the perihelion shift contains
two pure terms containing terms that depend on $l$ and $\lambda$
respectively along with some interfering terms. One of the pure
terms comes from the use of a bumblebee background and the other
appears because of the use of non-commutative spacetime proposed
by Snyder. The interfering terms that appear are very very small
since the term contains the product of two very small factors
$\lambda$ and $l$. Although the interfering term is negligibly
small, it has some theoretical consequence, e.g., symmetric
contribution may get disturbed by an asymmetric term - the reverse
perhaps does not get so. It is true that both the parameters
$\lambda$ and $l$ are small. A physical result containing some
arbitrary parameter wold be physically sensitive if the parameters
can be constrained from experimental observation. Both the
parameters can be put under constraints using the same arguments
available in \cite{MIGGRAV} and \cite{CASANA}. The conjecture used
there is that the correction would be of the order of general
relativistic correction. Here, we can, indeed, put a restriction
on the parameters $\lambda$ and $l$ following the same argument.
The upper bound of the predicted values of the parameters is given
in $TABLE-1$. Note that the predicted values of $l$ are identical
to the values obtained in the article \cite{CASANA} values. It is
indeed the expected result. Our study enables us to put another
constraint the parameter $\lambda$ when the constrained on $l$ is
known if it is taken into consideration that the correction that
appeared in the form of interference which is proportional to
$\lambda l$ would be of the same order of magnitude of the pure
terms. Note that the first-order correction term is proportional
to $\lambda l$. So if we are intended to constrain the interfering
term it will provide new information that the parameters may have
reciprocal out-turn between each other. When the contribution
through one parameter dominates, the contribution from the other
will get suppressed automatically.
\section{Acknowledgement}
AR would like to acknowledge the facilities extended to him during
his visit to the I.U.C.A.A, Pune. He would also like to thank the
Director of Saha Institute of Nuclear Physics, Kolkata, for
providing library facilities of the Institute.

\end{document}